\documentclass[aps,prx,twocolumn,superscriptaddress,amssymb]{revtex4}
\usepackage{tabularx}
\usepackage{bm}
\usepackage{euscript}
\usepackage{epsfig,psfrag,subfigure}
\usepackage{graphicx}
\usepackage{color}
\usepackage{amsfonts}
\usepackage{exscale}
\usepackage{amsbsy}
\usepackage{multirow}

\def\avg#1{\left\langle#1\right\rangle}

\def\be{\begin{equation}}       \def\ee{\end{equation}}
\def\bea{\begin{eqnarray}}      \def\eea{\end{eqnarray}}
\def\ba{\begin{array} }
\def\ea{\end{array} }

\def\nn{\nonumber}

\def\=>{\Rightarrow}
\def\>{\rightarrow}

\def\Fig#1{Fig.~\ref{#1}}

\def\Tr{\mathrm{Tr}}

\renewcommand{\v}[1]{{\bf #1}}

\renewcommand{\>}{\rangle}

\begin{document}

\title{Field-induced quantum spin liquid in the Kitaev-Heisenberg model and its relation to $\alpha$-RuCl$_3$}
\author{Yi-Fan Jiang}
\affiliation{Stanford Institute for Materials and Energy Sciences, SLAC National Accelerator Laboratory and Stanford University, Menlo Park, CA 94025, USA}
\author{Thomas P. Devereaux}
\affiliation{Stanford Institute for Materials and Energy Sciences, SLAC National Accelerator Laboratory and Stanford University, Menlo Park, CA 94025, USA}
\affiliation{{Department of Materials Science and Engineering, Stanford University, Stanford, CA 94305, USA}}
\author{Hong-Chen Jiang}
\email{hcjiang@stanford.edu }
\affiliation{Stanford Institute for Materials and Energy Sciences, SLAC National Accelerator Laboratory and Stanford University, Menlo Park, CA 94025, USA}

\begin{abstract}
Recently considerable excitement has arisen due to the experimental observation of a field-induced spin liquid phase in the compound $\alpha$-RuCl$_3$. However, the nature of this putative spin liquid phase and the relevant microscopic model Hamiltonian remain still unclear. In this work, we address these questions by performing large-scale numerical simulations of a generalized Kitaev-Heisenberg model proposed to describe the physics of $\alpha$-RuCl$_3$. While there is no evidence for an intermediate phase for in-plane magnetic fields, our results strongly suggest that a stable intermediate spin liquid phase, sandwiched between a magnetically ordered phase at low fields and a high-field polarized phase, can be induced by out-of-plane magnetic fields. Moreover, we show that this field-induced spin liquid phase can be smoothly connected to a spin liquid possessing a spinon Fermi surface as proposed recently for the Kitaev model. The relevance of our results to $\alpha$-RuCl$_3$ is also discussed.
\end{abstract}
\date{\today}
\maketitle

The search for quantum spin liquids (QSLs) in frustrated quantum magnets has enjoyed a
surge of interest in modern condensed matter physics\cite{Anderson1973,Balents2010}. The Kitaev model on the honeycomb lattice is exact solvable and known to exhibit a gapless spin liquid ground state (for equal coupling along the links) that can be gapped out into a topological phase with non-Abelian quasiparticle excitations by certain time-reversal symmetry breaking perturbations such as magnetic fields\cite{Kitaev2006,Jiang2011,Sela2014,Balents2014,Schaffer2016,Zhu2018,Hickey2018,Jiang2018,Zou2018,Patel2018}, which is an important ingredient for fault-tolerant quantum computation\cite{Nayak2008}. Consequently, there has been enormous interest in exploring the possible realization of Kitaev physics in a large family of layered Mott insulators with strong spin-orbit couplings such as $\alpha$-RuCl$_3$\cite{Jackeli2009,Singh2010,Plumb2014,Rau2016,Yadav2018}. Although, the material exhibits a ``zigzag'' long-range magnetic order below $T_N=7\sim 14$ K, recent experiments show that a moderate external magnetic field, around 8T, can suppress the order and drive $\alpha$-RuCl$_3$ into a paramagnetic phase, which is a plausible candidate for field-induced QSLs\cite{Wolter2017, Leahy2017, Baek2017, Sears2017, Zheng2017, Yu2018,Hentrich2018,Kasahara2018}. However, the nature of this putative spin liquid phase, either gapless or gapped is still under intense debate. Meanwhile, the details of the larger phase diagram under magnetic field remains largely unknown as well.

Theoretically, distinct models in the context of Eq.(\ref{Eq:Ham}) (given below) were proposed to understand the experimental results\cite{Banerjee2016,Banerjee2017,Winter2017,Winter2018,Ran2017,Hou2017,Kim2015,Kim2016,Winter2016,
Wang2017,Yadav2016,Gohlke2018}. In addition to the Kitaev interaction, a variety of other interactions, including the Heisenberg and off-diagonal spin-orbit coupling $\Gamma$-term, are proposed to be necessary to understand the experimental results of $\alpha$-RuCl$_3$. However, which one is correct is still unclear. For instance,  neutron scattering measurements\cite{Banerjee2016,Banerjee2017} suggest that an antiferromagnetic (AFM) Kitaev coupling is necessary, which is in sharp contrast to some other studies\cite{Wang2017,Winter2016,Kim2016} which suggest that a ferromagnetic (FM) Kitaev coupling is necessary. Moreover, the relative strength of various interactions varies greatly from study to study, raising additional difficulties in understanding the experimental results of $\alpha$-RuCl$_3$.

To answer these questions, we determine the phase diagram of the model in Eq.(\ref{Eq:Ham}) under magnetic fields using exact diagonalization (ED) and density matrix renormalization group (DMRG)\cite{White1992} and search for the appropriate set of parameters for $\alpha$-RuCl$_3$. To be consistent with experiments, we start with an appropriate set of couplings so that the ground state of the model in the absence of magnetic field hosts zigzag long-range magnetic order. Subsequently, we explore the entire phase diagram of the model with either in-plane or out-of-plane magnetic fields to search for signatures of putative field-induced paramagnetic phases, before entering a fully polarized state for sufficiently strong magnetic fields. Our study indicates that for most sets of parameters extracted from previous studies, a field-induced spin liquid or an intermediate phase between a zigzag ordered phase at low magnetic field and a fully polarized phase at high magnetic field, is absent. However, for the set of parameters with AFM Kitaev coupling and out-of-plane magnetic fields, we find strong evidence of an intermediate field-induced paramagnetic phase consistent with a gapless spin liquid with a spinon Fermi surface as proposed recently\cite{Hickey2018,Jiang2018,Zou2018,Patel2018}.

\begin{table}[t]
\centering
\begin{tabular}{| c | c | c | c | c | c | c | c | c | c | c |}
\hline
Set & $K_1$ & $\Gamma$ & $J_1$ & $J_2$ & $J_3$ & $K_3$ & Pattern & \multicolumn{2} {m{1.5cm}|} {\centering NIP} & Ref \\
\cline{9-10}
 &  &  &  &  &  &  &  & \hspace{1mm}$h_b$ \hspace{1mm} & $h_{c^*}$ &  \\
\hline
1 & 7 &  & -4.6 &  &  &  & Zigzag & 0 & 1 & \cite{Banerjee2017} \\
2 & -5 & 2.5 & -0.5 &  & 0.5 &  & Zigzag & 0 & 0 & \cite{Winter2017} \\
3 & -10.6 & 3.8 & -1.8 &  & 1.25 & 0.65 & Zigzag & 0 & 0 & \cite{Hou2017} \\
4 & -6.8 & 9.5 &  &  &  &  & ID & 0 & 0 & \cite{Ran2017} \\
5 & -5.5 & 7.6 &  &  &  &  & ID & 0 & 0 & \cite{Kim2015} \\
6 & -8 & 4 & -1 &  &  &  & ID & 0 & 0 & \cite{Kim2016} \\
7 & -6.6 & 6.6 & -1.7 &  & 2.7 &  & Zigzag & 0 & 0 & \cite{Winter2016} \\
8 & 17 & 12 & -12 &  &  &  & 120$^0$ & 1 & 0 & \cite{Wang2017} \\
9 & -5.6 & -1 & 1.2 & 0.3 & 0.3 &  & 120$^0$ & 1 & 0 & \cite{Yadav2016} \\
\hline
\end{tabular}
\caption{Summary of numerical results for finite magnetic fields $\v{h_b}$ and $\v{h_{c^*}}$ for various sets of parameter extracted from Ref.\cite{Banerjee2017,Winter2017,Winter2018,Hou2017,Ran2017,Kim2015,Kim2016,Winter2016,Wang2017,Yadav2016}. ``Pattern" labels the pattern of magnetic ordering in the ground state at low magnetic fields, and ``NIP" denotes the number of intermediate phases between low-field and high-field polarized phases.}
\label{Table1}
\end{table}

{\bf Model Hamiltonian:} %
We study the following generalized Kitaev-Heisenberg model proposed in previous studies \cite{Banerjee2016,Banerjee2017,Winter2017,Winter2018,Ran2017,Kim2015,Kim2016,Winter2016,Wang2017,Yadav2016}, which is defined by the Hamiltonian%
\begin{eqnarray} \label{Eq:Ham}
H&=&\sum_{\avg{ij}} J_1 \vec{S}_i\cdot\vec{S}_j + K_1 S_i^\gamma S_j^\gamma + \Gamma( S_i^\alpha S_j^\beta +S_i^\beta S_j^\alpha )\nn\\
&+&J_2\sum_{\langle\langle ij\rangle\rangle}\vec{S}_i\cdot\vec{S}_j+\sum_{\langle\langle\langle ij\rangle\rangle\rangle}J_3\vec{S}_i\cdot\vec{S}_j + K_3 S_i^\gamma S_j^\gamma \, .
\end{eqnarray}
Here $K_1$ and $K_3$ are the nearest-neighbor (NN) and third-neighbor Kitaev interactions, respectively, and $\Gamma$ is the NN off-diagonal spin-orbit coupling. $J_1$, $J_2$ and $J_3$ are the NN, second and third-neighbor Heisenberg interactions, respectively. $\{\alpha, \beta, \gamma \}$ determine the bond-dependent Kitaev and $\Gamma$ interaction. On the $z$ bond $\{\alpha, \beta, \gamma \}=\{ x,y,z \}$, and the form of $x$ and $y$ bonds are obtained by cyclic permutation.
In this paper, we will systematically investigate the ground state properties of this model by employing ED and DMRG methods.

\begin{figure}[t]
\centering
\includegraphics[width=\linewidth]{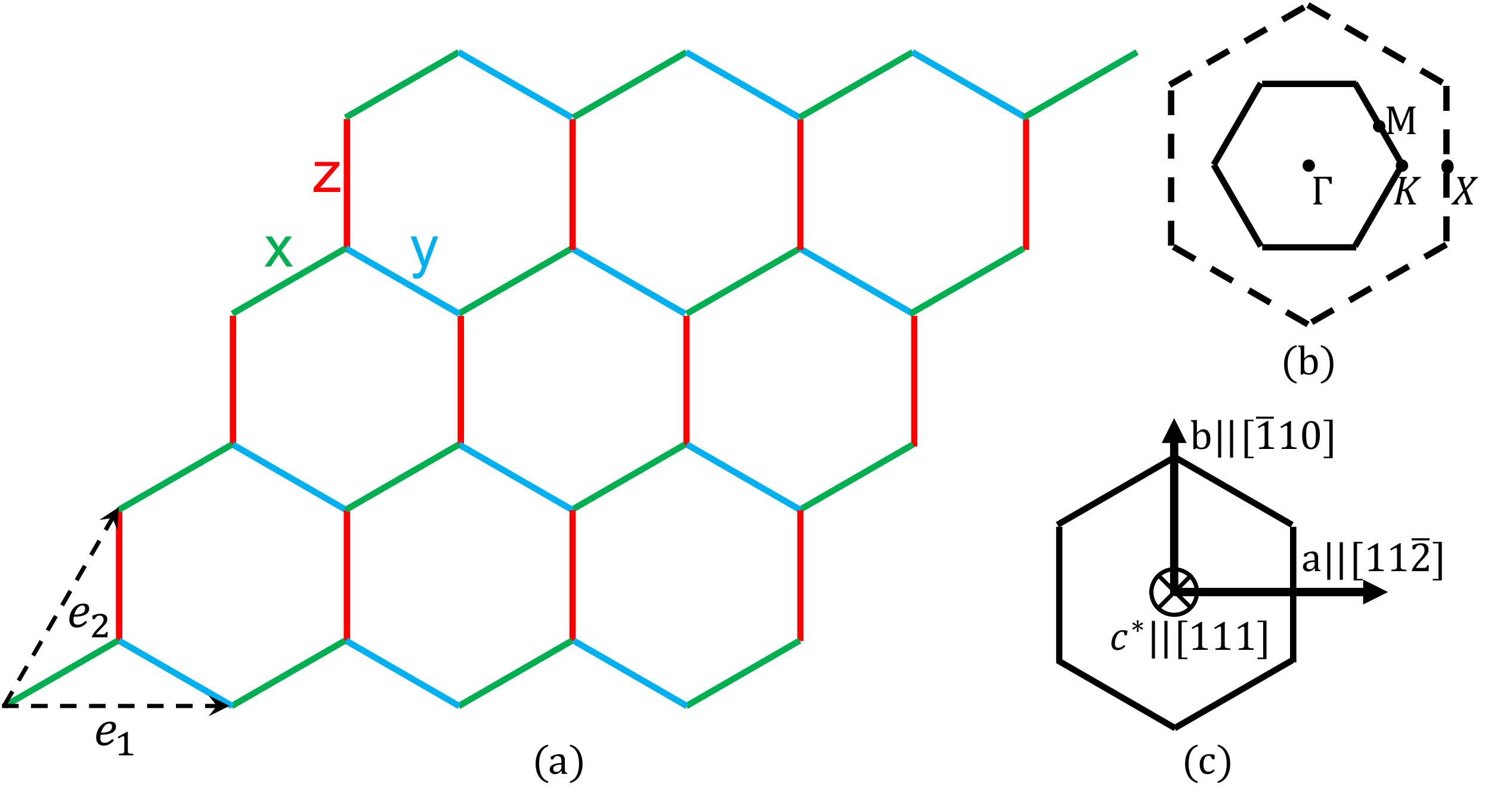}
\caption{(a) Honeycomb cylinder with periodic and open boundary conditions along the directions specified by the lattice basis vectors, $e_2$ and $e_1$, respectively. $L_x$ and $L_y$ are the number of unit cells in the $e_1$ and $e_2$ directions. Bond directions $\gamma=x,y,z$ are labeled by different colors. (b) The first and second Brillouin zone and high symmetry points. (c) Directions of the external magnetic field in spin space.}
\label{Fig:Lattice}
\end{figure}

The lattice geometry used in our simulations is depicted in Fig. \ref{Fig:Lattice}, where $e_1$=($\sqrt{3}$,0) and $e_2$=($\frac{\sqrt{3}}{2},\frac{3}{2}$) denote the two basis vectors. We consider honeycomb cylinders with periodic (open) boundary conditions in the $e_2$ ($e_1$) direction. Here, we focus on cylinders with width $L_y$ and length $L_x$, where $L_y$ and $L_x$ are the number of unit cells ($2L_y$ and $2L_x$ are the number of sites) along the $e_2$ and $e_1$ directions, respectively. The total number of sites is $N=2\times L_x\times L_y$. In this paper, we focus primarily on cylinders with width $L_y$=3 and $L_y$=4, and have also checked our results using different lattice geometries such as the $C_6$ rotationally symmetric $N=24$-site cluster illustrated in Supplemental Material (SM). 

\textbf{Principal results:} %
We have investigated the ground state properties of the model Hamiltonian in Eq.(\ref{Eq:Ham}) using the proposed sets of parameters summarized in Table \ref{Table1}. The parameter sets 1-4 denote the sets of  interactions extracted from neutron scattering experiments\cite{Banerjee2016, Banerjee2017, Winter2017,Winter2018,Ran2017, Hou2017}, and sets 5-8 are determined from density functional theory (DFT) \cite{Kim2015,Kim2016,Winter2016,Wang2017}, while set 9 is estimated by fitting the magnetization curve using quantum chemistry techniques\cite{Yadav2016}. The column ``Pattern'' denotes the pattern of the magnetic order in the ground state for corresponding set of parameters, which is determined by calculating the spin-spin correlation function and spin structure factor defined in Eq.(\ref{Eq:Sk}). As illustrated in Fig.\ref{Fig:Lattice}, the ``zigzag'' order is featured by sharp peaks in the structure factor at the $M$ points, while ``$120^0 $'' order is peaked at $K$ or $K^\prime$ points in the Brillouin zone of the honeycomb lattice. 

Among all sets of parameters in Table \ref{Table1}, our results suggest that only parameter sets 1, 2, 3 and 7 exhibit zigzag ordered ground states where the structure factor $S(\vec{q})$ is peaked at $M$ points in the Brillouin zone, consistent experiments. On the contrary, the structure factor $S(\vec{q})$ for parameter sets 8 and 9 show sharp peaks at $K$ points which imply a $120^0$ order instead of a zigzag order. For parameter sets 4, 5 and 6, $S(\vec{q})$ only has very broad peaks which are not located at any high symmetry points such as $M$ or $K$ points, so we refer them to as possible incommensurate or disordered (ID) phases. More details of the results are given in the Supplemental Material.

To make a direct connection to experiment, we have also studied the ground-state properties of the model in Eq.(\ref{Eq:Ham}) under external uniform magnetic field $\v{h}$ given by%
\begin{eqnarray}
H_h=- \v{h} \cdot \sum_i \v{S}_i \, .
\end{eqnarray}
As shown in \Fig{Fig:Lattice}, axes of external magnetic field $a$, $b$ and $c^*$ of $\alpha$-RuCl$_3$ correspond to $[11\bar{2}]$, $[\bar{1}10]$ and $[111]$ respectively, which are labeled by spin directions $[xyz]$. In the following, we consider both $\v{h_{c^*}} \| [111]$ and $\v{h_{b}} \| [\bar{1}10]$ cases. For each set of parameters, the field-induced phase diagrams are determined using ED on $N=24$-site cluster 
and DMRG on $L_y=3\sim 4$ cylinder.

We have calculated both the ground state energy and magnetization as well as their derivatives to search for possible field-induced intermediate phases between the low-field phase and the fully polarized phase in high magnetic fields. The number of intermediate phases (NIP) for each set of parameters under magnetic field $\v{h_{c^*}} \| [111]$ and $\v{h_{b}} \| [\bar{1}10]$ is listed in column ``NIP'' in Table \ref{Table1}. More details of the simulation and results are provided in the SM. Surprisingly, for all the sets of parameters which host a zigzag magnetic order in low fields, we find that only parameter set 1 establishes a field-induced intermediate phase, which is absent for all the other sets of parameters. Since the parameter set 1 only has AFM Kitaev coupling $K_1>0$ and FM Heisenberg interaction $J_1<0$, our results suggest that they are crucial for stabilizing a field-induced intermediate phase. In the following, we will focus on this so-called Kitaev-Heisenberg model with special attention to its phase diagram and the field-induced intermediate phase.

\begin{figure}[tb]
\centering
\includegraphics[width=\linewidth]{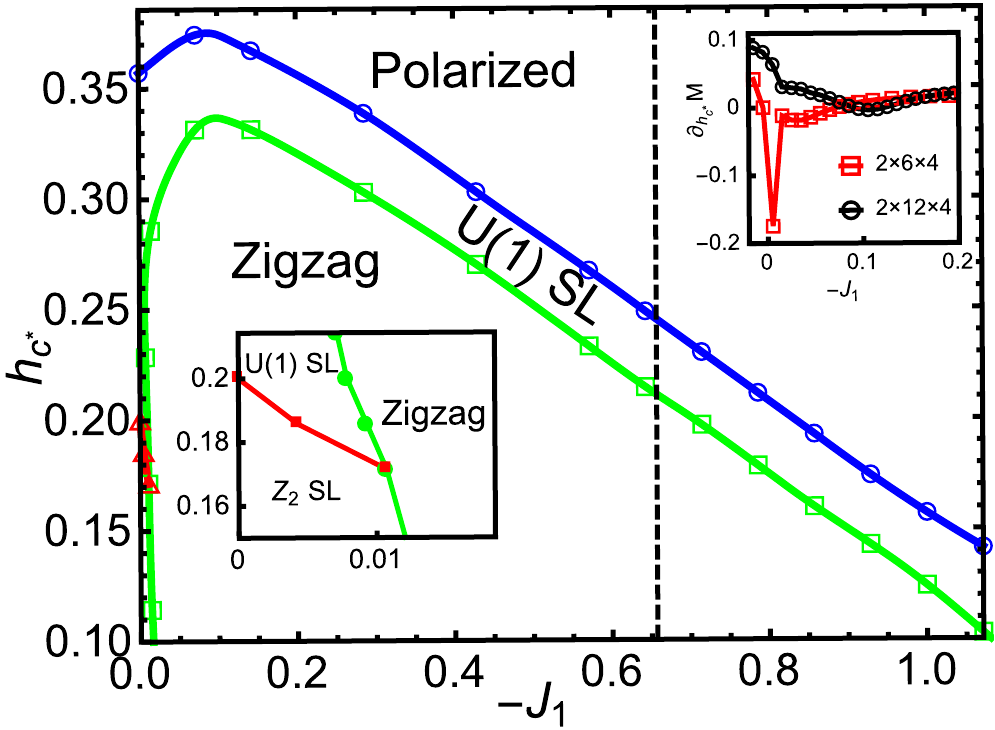}
\caption{Ground state phase diagram of the Kitaev-Heisenberg model on $L_y=4$ cylinder under magnetic field $\v{h_{c^*}}$. $Z_2$ and $U(1)$ SL denote the $Z_2$ gapped spin liquid and $U(1)$ gapless spin liquid with spinon Fermi surface, respectively. Insets: zoomed-in phase diagram with small $J_1$ (left) and derivative of magnetization at $h_{c^*}=0.343$ (right). Dashed line denotes the set of parameter extracted from neutron scattering experiments\cite{Banerjee2017}.}
\label{Fig:PhaseHc}
\end{figure}

\textbf{Kitaev-Heisenberg Model:} %
Previous studies have shown that the Kitaev-Heisenberg model has a rich phase diagram hosting distinct phases, including Neel, zigzag and spin liquid phases\cite{Chaloupka2013,Iregui2014,Gohlke2017,Gotfryd2017}. In particular, zigzag order can be induced by AFM Kitaev ($K_1>0$) and FM Heisenberg couplings ($J_1<0$). For small $J_1$ and zero magnetic field, it has been shown that a tiny $J_1/K_1 \sim -0.012$ is enough to drive the system into a zigzag ordered phase\cite{Gotfryd2017}. A natural question is how stable is the zigzag order against external magnetic fields and what is the nature of the possible subsequent phases. We will try to answer these questions for magnetic fields $\v{h_{c^*}} \| [111]$ and $\v{h_b} \| [\bar{1}10]$ separately. For simplicity, here we set $K_1=1$ as an energy unit.

{\bf $\v{h_{c^*}} \| [111]$: } The phase diagram of the Kitaev-Heisenberg model under out-of-plane magnetic fields $\v{h_{c^*}}$ is shown in \Fig{Fig:PhaseHc}. In the Kitaev limit, i.e., $J_1$=0, the system stays in a stable gapped spin liquid phase hosting non-Abelian Ising anyons until the magnetic field is higher than $h_{c^*} \sim 0.20$\cite{Zhu2018}. The system enters into a gapless spin liquid phase for higher magnetic fields, as shown in the left inset of \Fig{Fig:PhaseHc}, which is consistent with the spin liquid with spinon Fermi surfaces proposed recently\cite{Hickey2018,Jiang2018,Zou2018,Patel2018}. Finally, the system becomes fully polarized for sufficiently high magnetic fields $h_{c^*} > 0.36$.

For small $h_{c^*}$, we find that the zigzag ordered phase can be stabilized by $J_1$ interactions which occupies a big portion of the phase diagram in Fig.\ref{Fig:PhaseHc}.
Interestingly, for fairly high $h_{c^*}$, an intermediate phase is observed in a large range of $J_1$ including $J_1=-4.6/7\approx -0.657$ which is extracted from neutron scattering experiment of $\alpha$-RuCl$_3$\cite{Banerjee2016,Banerjee2017}. The phase boundaries among the zigzag, intermediate and polarized phases are determined by the derivatives of the ground state energy and the magnetization $|\v{m}|=\frac{1}{N}|\sum_i \avg{\v{S}_i}|$ as a function of $h_{c^*}$ and $J_1$. We have obtained consistent results on various different lattices, including $L_y=3$ and $L_y=4$ cylinders, and the $C_6$ rotationally symmetric $N=24$-site clusters which are consistent with recent study\cite{Hickey2018}. As an example, the derivative of magnetization at $h_{c^*}=0.343$ is shown in \Fig{Fig:PhaseHc} as the right inset. With the increase of system size, it is clear that the intermediate phase at large $J_1$ is smoothly connected to the gapless spin liquid phase of the Kitaev model under external magnetic field, indicating that the intermediate state is consistent with the gapless spin liquid with spinon Fermi surface proposed recently\cite{Hickey2018,Jiang2018,Zou2018,Patel2018}. More details are provided in the SM.

\begin{figure}[tb]
\centering
\includegraphics[width=\linewidth]{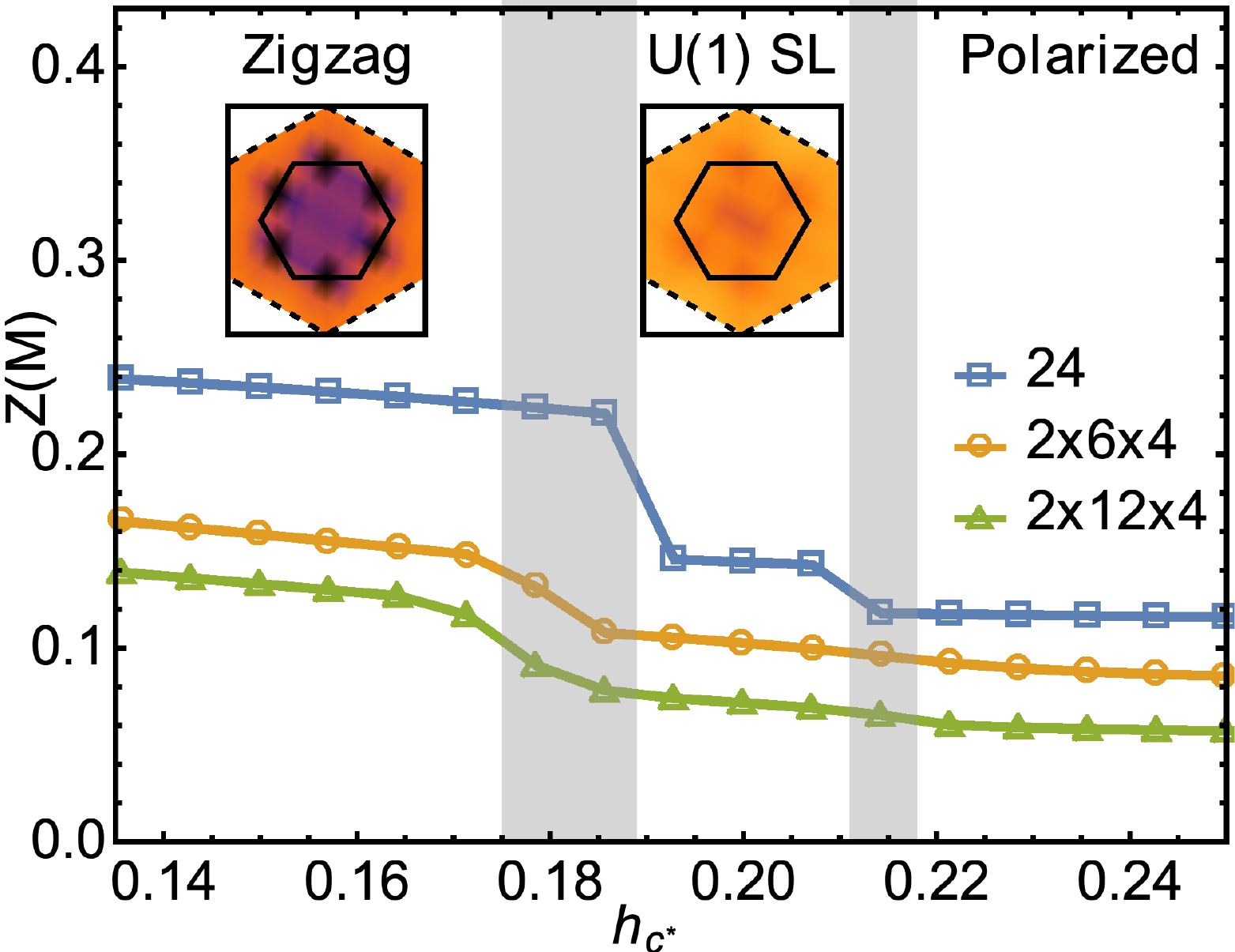}
\caption{Zigzag order parameter $Z(M)=\frac{1}{3}\sum_{i=1}^3 \sqrt{S(\v{M}_i)/N}$ along the dashed line in Fig.\ref{Fig:PhaseHc} as a function of $h_{c^*}$. Phase transitions between distinct phases are labelled by shaded regions. Inset: examples of the spin structure factor $S(\v{q})$ in the zigzag and spin liquid phases on $L_y=4$ cylinder of length $L_x=6$.
}
\label{Fig:SkHc}
\end{figure}

To determine the precise nature of the intermediate phase between the zigzag and fully polarized phases, we have calculated the spin structure factor defined as %
\bea
S(\v{q}) = \frac{1}{N}\sum_{ij} \avg{(\v{S}_i - \v{m}) \cdot (\v{S}_j - \v{m})}e^{i \v{q} \cdot (\v{r}_i-\v{r}_j)} ,
\label{Eq:Sk}
\eea
As shown in the inset of Fig. \ref{Fig:SkHc}, $S(\v{q})$ shows sharp peaks at different $M$ points in the first Brillouin zone, which is a clear feature of zigzag order. On the contrary, $S(\v{q})$ is almost featureless in the intermediate phase, indicating the absence of any magnetic order. We further define the zigzag order parameter as $Z(M)=\frac{1}{3}\sum_{i=1}^3 \sqrt{S(\v{M}_i)/N}$, where $S(\v{M}_i)$ is the spin structure factor at three distinct $M$ points in the first Brillouin zone. The phase boundaries are then located by the peak positions of the derivatives of the zigzag order parameter $\partial Z(M)/\partial h$ (shown in the SM).
 
\begin{figure}[tb]
\centering
\includegraphics[width=\linewidth]{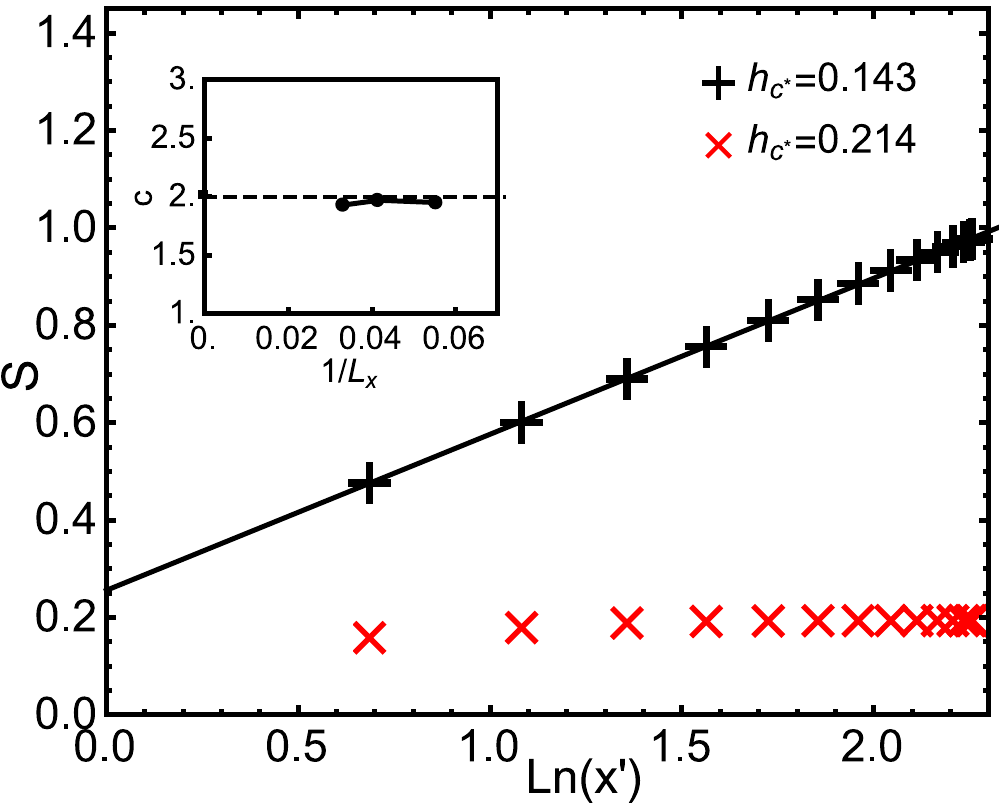}
\caption{(Color online) Von Neumann entanglement entropy $S$ on $L_y=4$ cylinder of length $L_x=30$ for the $J_1=-1.0$ model, where $x^\prime=\frac{L_x}{\pi}\sin(\frac{\pi x}{L_x})$. The extracted central charge $c$ is shown in the inset where the dashed line is a guide for eyes.}
\label{Fig:EE}
\end{figure}

To further characterize the intermediate phase, we have also calculated the von Neumann entropy $S=-\Tr\rho \log \rho$ using DMRG, where $\rho$ is the reduce density matrix of subsystem with length $x$. For a $1+1$ dimensional critical system described by a conformal field theory (CFT), it is known that $S(x) =\frac{c}{6}\ln\big[\frac{L_x}{\pi}\sin(\frac{\pi x}{L_x})\big]+\tilde{c}$ on a cylinder of length $L_x$, where $c$ is the central charge of the CFT and $\tilde{c}$ is a model-dependent constant. Using this formula we extracted the central charge $c$ numerically for cylinders of width $L_y=4$ and length $L_x=30$, as shown in Fig.\ref{Fig:EE}. Here we keep up to $m=1200$ block states with a truncation error $\epsilon\leq 10^{-8}$. We choose two representative strengths of magnetic field $h_{c^*}=0.143$ and $0.214$ deep inside the intermediate and polarized phases, respectively. The $h_{c^*}=0.143$ data has $c\sim1.94$ suggesting that there are $c=2$ gapless modes, which is consistent with the gapless spin liquid with spinon Fermi surfaces\cite{Hickey2018,Jiang2018,Zou2018,Patel2018}. On the contrary, $c\sim 0$ at $h_{c^*}=0.214$ in the fully polarized phase, indicating a gapped ground state. Moreover, we also calculate the central charge for cylinders of different lengths $L_x=$18, 24 with magnetic field $h_{c^*}=0.143$. The result is shown in the inset of Fig.\ref{Fig:EE} which is consistent with $L_x=30$ data where finite-size effects are negligible.

\begin{figure}[t]
\centering
\includegraphics[width=\linewidth]{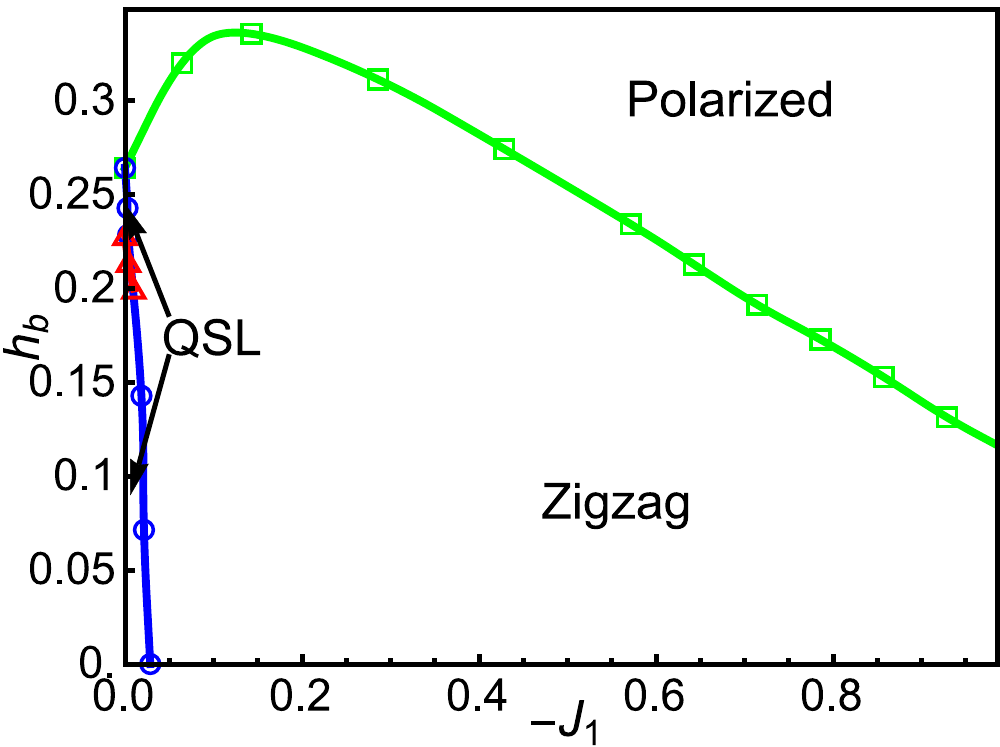}
\caption{(Color online)  Ground state phase diagram of the Kitaev-Heisenberg model on $L_y=4$ cylinder under magnetic field $\v{h_{b}}$.}
\label{Fig:PhaseHb}
\end{figure}

{\bf $\v{h_b} \| [\bar{1}10]$:} %
To make a more thorough connection with experiments, we have further calculated the ground state phase diagram of the Kitaev-Heisenberg model under magnetic fields $\v{h_b}$ in \Fig{Fig:PhaseHb}. For small $\v{h_b}$, the phase diagram is very similar with the one of $\v{h_{c^*}}$, which has the spin liquid phase for small $J_1$ while zigzag ordered phase for larger $J_1$. In the Kitaev limit, i.e., $J_1$=0, an intermediate spin liquid phase is also present between the Kitaev spin liquid at low fields and the fully polarized phase at high fields, with phase boundaries located at $h_{b} \sim 0.22$ and $0.27$ respectively. However, contrary to the $\v{h_{c^*}}$ case, the gapless spin liquid phase is not very stable against $J_1$ and the system quickly enters into either a fully polarized phase or a zigzag ordered phase depending on the strength of the magnetic field, and there is no field-induced intermediate or spin liquid phases.

{\bf Conclusion and outlook: } %
In this work, we have systematically studied the ground state properties of the model Hamiltonian which has been proposed to understand the physics of $\alpha$-RuCl$_3$. Our results suggest that among the distinct types of interactions, the AFM Kitatev and FM Heisenberg interactions are crucial to determining the phase diagram under external magnetic fields. While there is no evidence for an intermediate paramagnetic phase under in-plane magnetic fields $\v{h_b}$, our results strongly suggest that a field-induced spin liquid phase can be achieved by applying out-of-plane magnetic fields $\v{h}_{c^*}$. 
However, this seems partially contrary to experiments, which instead suggest that a putative intermediate phase can be realized under field oriented both in $ab$ plane\cite{Wolter2017,Leahy2017,Sears2017,Zheng2017,Yu2018} and $60^0$ off $ab$ plane\cite{Baek2017}. This might be partially attributed to the $g$-factor that we have used in our calculation, where we have ignored the anisotropy for simplicity. However, if we extract the critical values of magnetic field at phase boundaries along the dashed line in Fig.\ref{Fig:PhaseHc}, which corresponds to the set of parameters used in neutron scattering experiments\cite{Banerjee2017}, we obtain $\mu_0 H_{c1}\sim \frac{25}{g} T$ and $\mu_0 H_{c2}\sim \frac{30}{g} T$. If we assume the electron spin $g$-factor $g \sim 2$, the critical magnetic field $H_{c1} $ will be consistent with the NMR measurement\cite{Baek2017}, suggesting that the $g$-factor may not be the major reason for this discrepancy. This raises the possibility that some ingredients which are crucial to obtain a field-induced paramagnetic phase under $\v{h}_{b}$ may be missing in the proposed model Hamiltonian in Eq.(\ref{Eq:Ham}), which will be an important question to be investigated in the future work.

{\it Acknowledgment: } %
We acknowledge Yuan-Ming Lu, Simon Trebst and Ciarán Hickey for insightful discussions and suggestions on our manuscript. This work was supported by the Department of Energy, Office of Science, Basic Energy Sciences, Materials Sciences and Engineering Division under Contract DE-AC02-76SF00515. Parts of the computing for this project was performed on the Sherlock cluster.

\renewcommand{\theequation}{A\arabic{equation}}
\setcounter{equation}{0}
\renewcommand{\thefigure}{A\arabic{figure}}
\setcounter{figure}{0}
\renewcommand{\thetable}{A\arabic{table}}
\setcounter{table}{0}
\begin{widetext}
\section{Supplemental Material}
\subsection{Numerical setup} 
We employ both ED and DMRG simulations to investigate the field induced phases of the models proposed for $\alpha$-RuCl$_3$. In the main text, most of the simulation of Kitaev-Heisenberg model are based on the $2\times6\times4$ honeycomb cylinder. Due to the lacking of spin rotational symmetry, the number of DMRG block states is limited. Here we keep up to $m=1200$ DMRG block states, which gives truncation errors smaller than $10^{-8}$. For the rest models listed in Table \ref{Table1}, we similarly determine their ground-state properties by DMRG calculations on $L_y=3\sim 4$ cylinders and their field induced phase phenomena by ED calculation on the 24-site PBC cluster illustrated in \Fig{Fig:24site}. 

\begin{figure}[hbt]
\centering
\includegraphics[height=5.0cm]{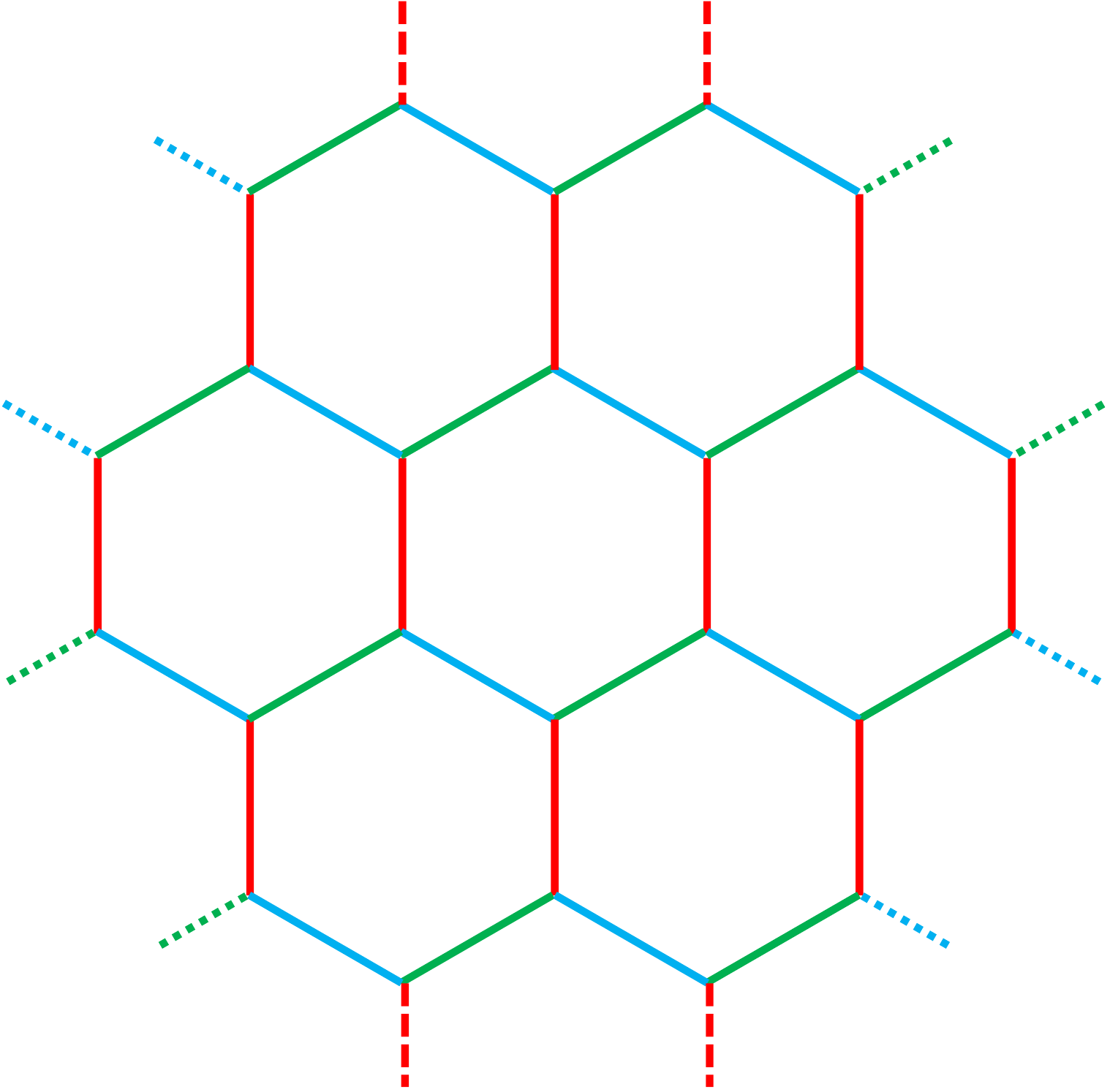}
\caption{$C_6$ rotationally symmetric 24-site cluster with periodic boundary conditions. Bond directions $\gamma=x,y,z$ are labelled by different colors.}
\label{Fig:24site}
\end{figure}

\begin{figure}[hbt]
\centering
\includegraphics[width=\linewidth]{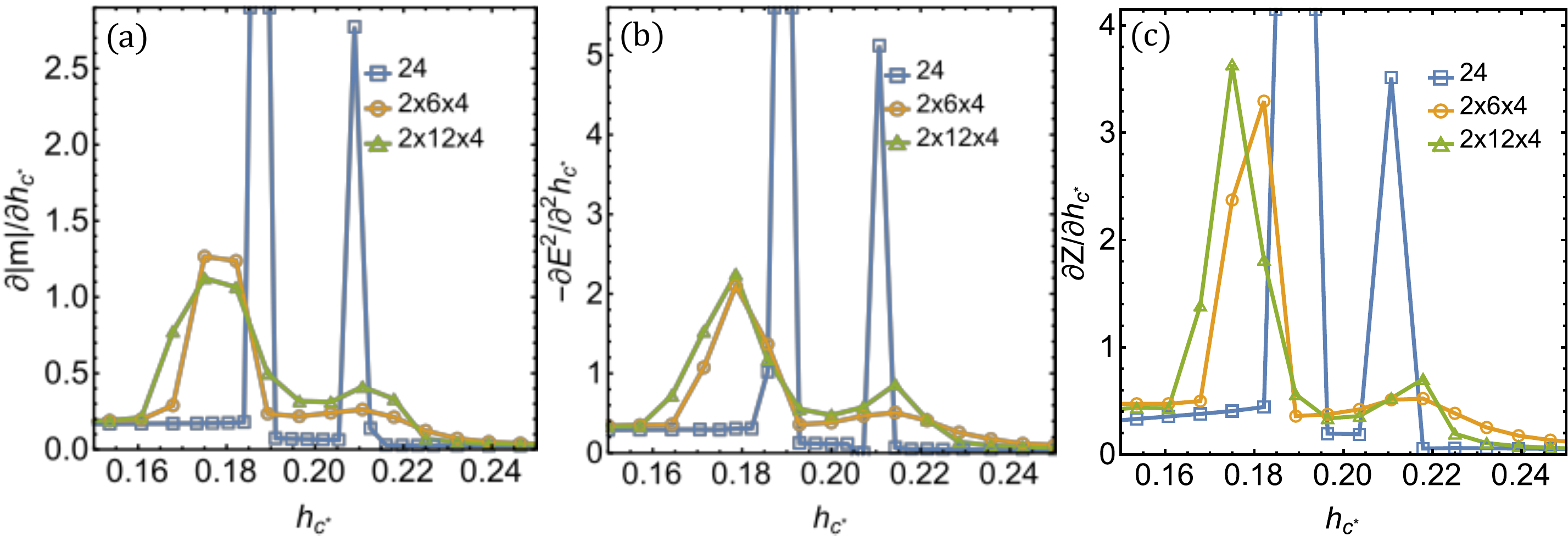}
\caption{(a) The derivative of magnetization along the dashed line in Fig.2 as a function of field $\v(h_{c^*})$. (b) The second derivative of total energy of the same model. (c) The derivative of zigzag order parameter defined in main text.}
\label{Fig:set1_KH}
\end{figure}

\subsection{Intermediate phase in Kitaev-Heisenberg model}
Here we provide more evidence of the two phase transitions exhibited in Fig.\ref{Fig:SkHc} ($J_1= - 4.6/7$ point). We calculate the derivative of magnetization, energy and zigzag order parameter along the dashed line in Fig. \ref{Fig:PhaseHc} on three different lattice geometry. As shown in Fig. \ref{Fig:set1_KH}, all of the derivatives support the phase transitions discussed in the main text.

\subsection{Spin structure factor}
We calculate the spin structure factor of the zero-field ground state of models 2-9 listed in Table \ref{Table1}. As illustrated in \Fig{Fig:Lattice} (b) in the main text, the peaks of spin structure factors at M, X and K points corresponding to zigzag, stripy and 120$^\circ$ magnetic order respectively. The DMRG simulation of most of the models are based on the $2\times6\times4$ cylinder, but for model 8 and 9 which exhibit 120$^\circ$ orders, we restrict the calculation to the $L_y=3$ cylinders to avoid frustration of the magnetic order. The result of models 2-9 are shown in \Fig{Fig:struct} (a-h) respectively, we can clearly see that only (a), (b) and (f) exhibit strong zigzag order, which correspond to model 2, 3 and 7 of Table \ref{Table1} in main test. 

\begin{figure}[hbt]
\centering
\includegraphics[width=\linewidth]{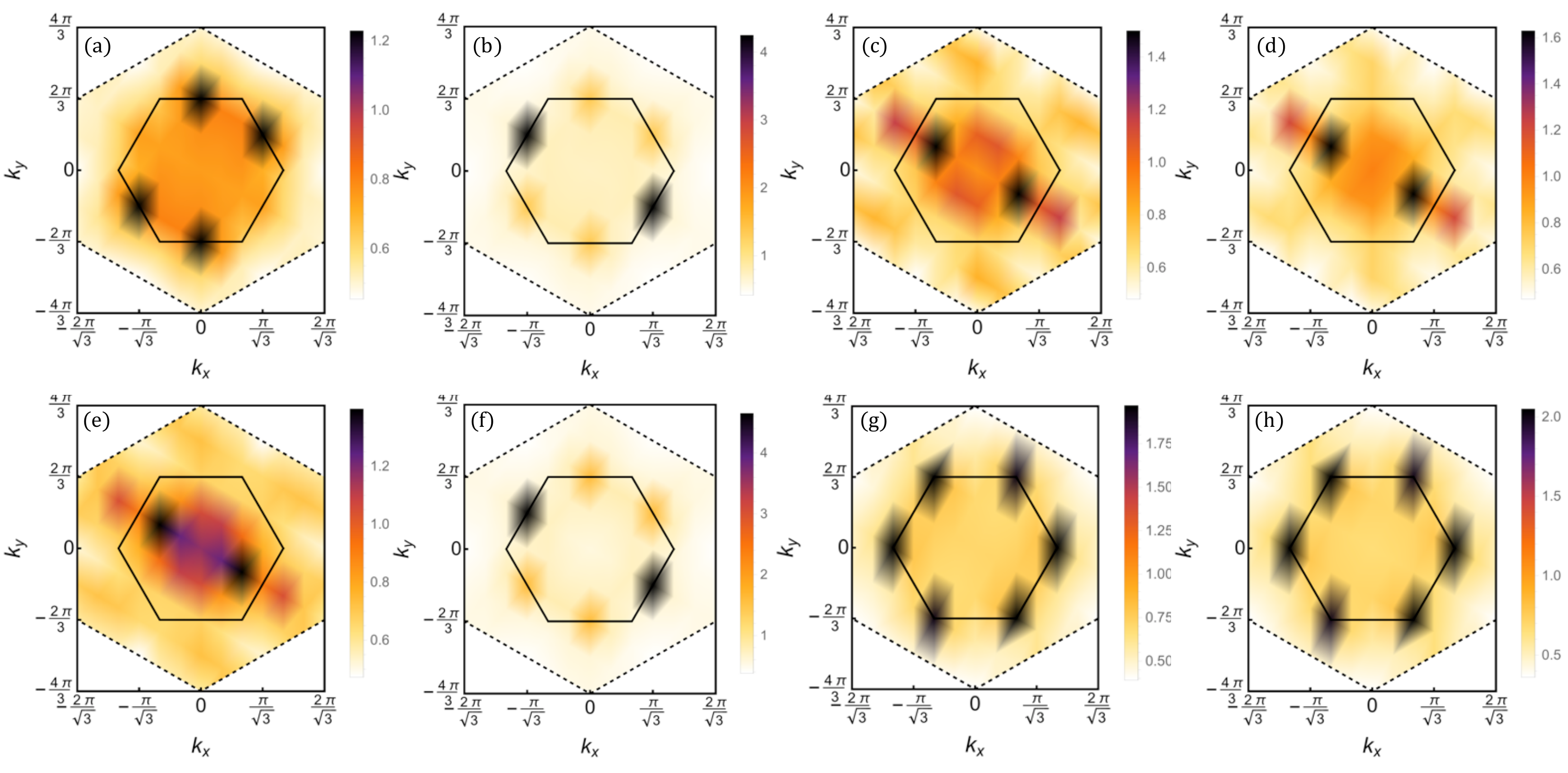}
\caption{Spin structure factors of model 2-9 listed in Table \ref{Table1} in main text. Fig (g) and (h) are the results from $2\times6\times3$ cylinder. The rest of the results are calculated on $2\times6\times4$ cylinder. Among model 2-9, only set 2, 3 and 7 exhibit sharp peaks at M points in the first Brillouin zone.}
\label{Fig:struct}
\end{figure}

\subsection{Derivative of the magnetization and ground state energy}
In Fig. \ref{Fig:dmag} and Fig. \ref{Fig:dmag}, we show the derivative of the magnetization and the second derivative of ground state energy as a function of the external magnetic field to determine the possible phase transitions of model 2-9 in Table \ref{Table1}. Both $\v{h_{c^*}} \| [111]$ and $\v{h_b} \| [\bar{1}10]$ fields are considered. The magnetization and energy is obtained by ED calculation on 24-site PBC cluster illustrated in \Fig{Fig:24site}. The amplitude of the external fields $h$ are pushed to at least 2.5 such that the magnetizations are nearly saturated for each model. For model 2, 3 and 7 which host long-range zigzag order at zero-field limit, we do not observe any intermediate phase in (a), (b) and (f) of \Fig{Fig:d2e} and \Fig{Fig:dmag}.

\begin{figure}[hbt]
\centering
\includegraphics[width=\linewidth]{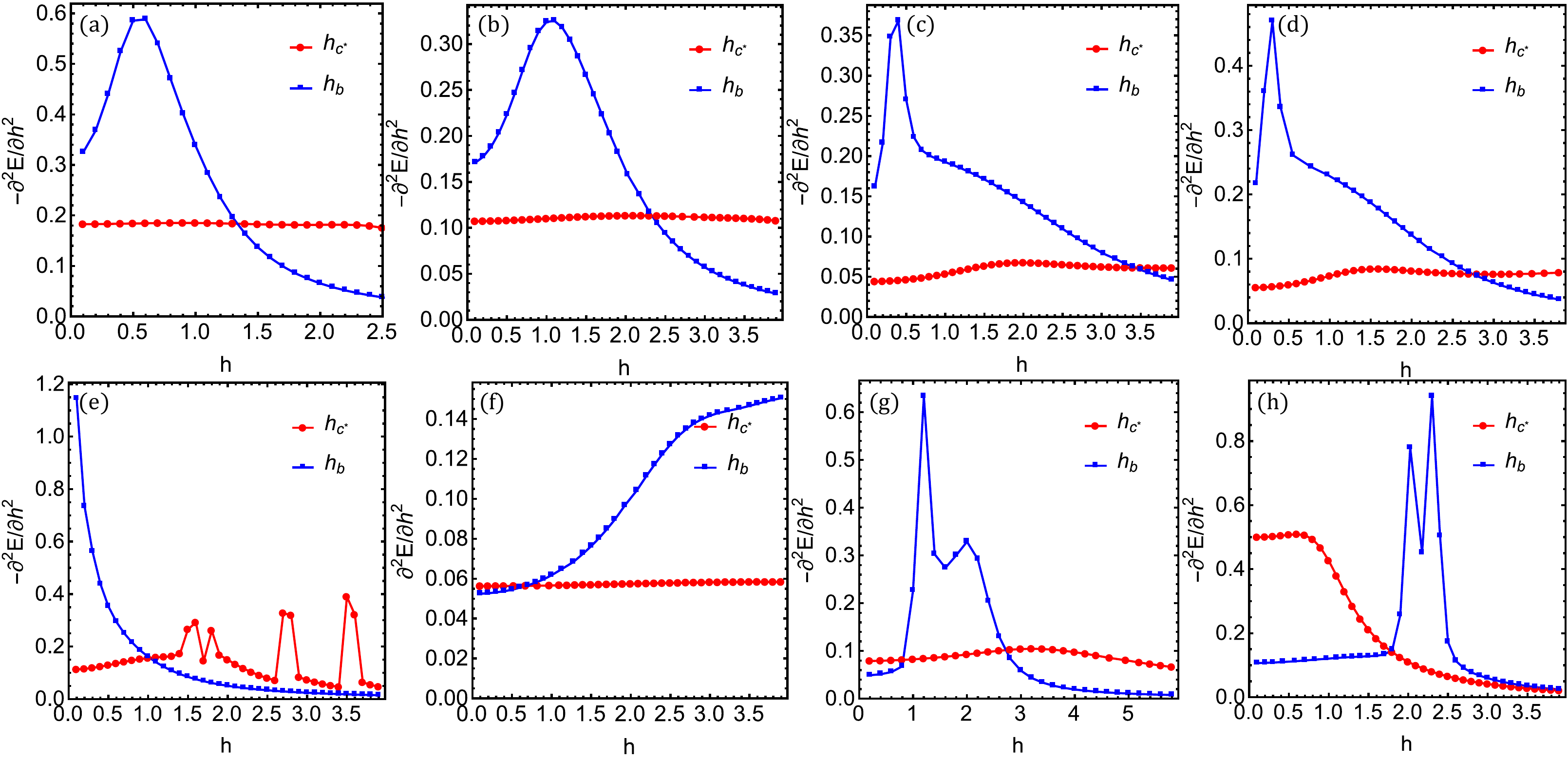}
\caption{The second derivative of total energy as a function of amplitude of field $h$. Red and blue curves stand for $\v{h_{c^*}} \| [111]$ and $\v{h_b} \| [\bar{1}10]$ fields respectively. The results are obtained by ED calculation on 24-site PBC cluster.}
\label{Fig:d2e}
\end{figure}

\begin{figure}[hbt]
\centering
\includegraphics[width=\linewidth]{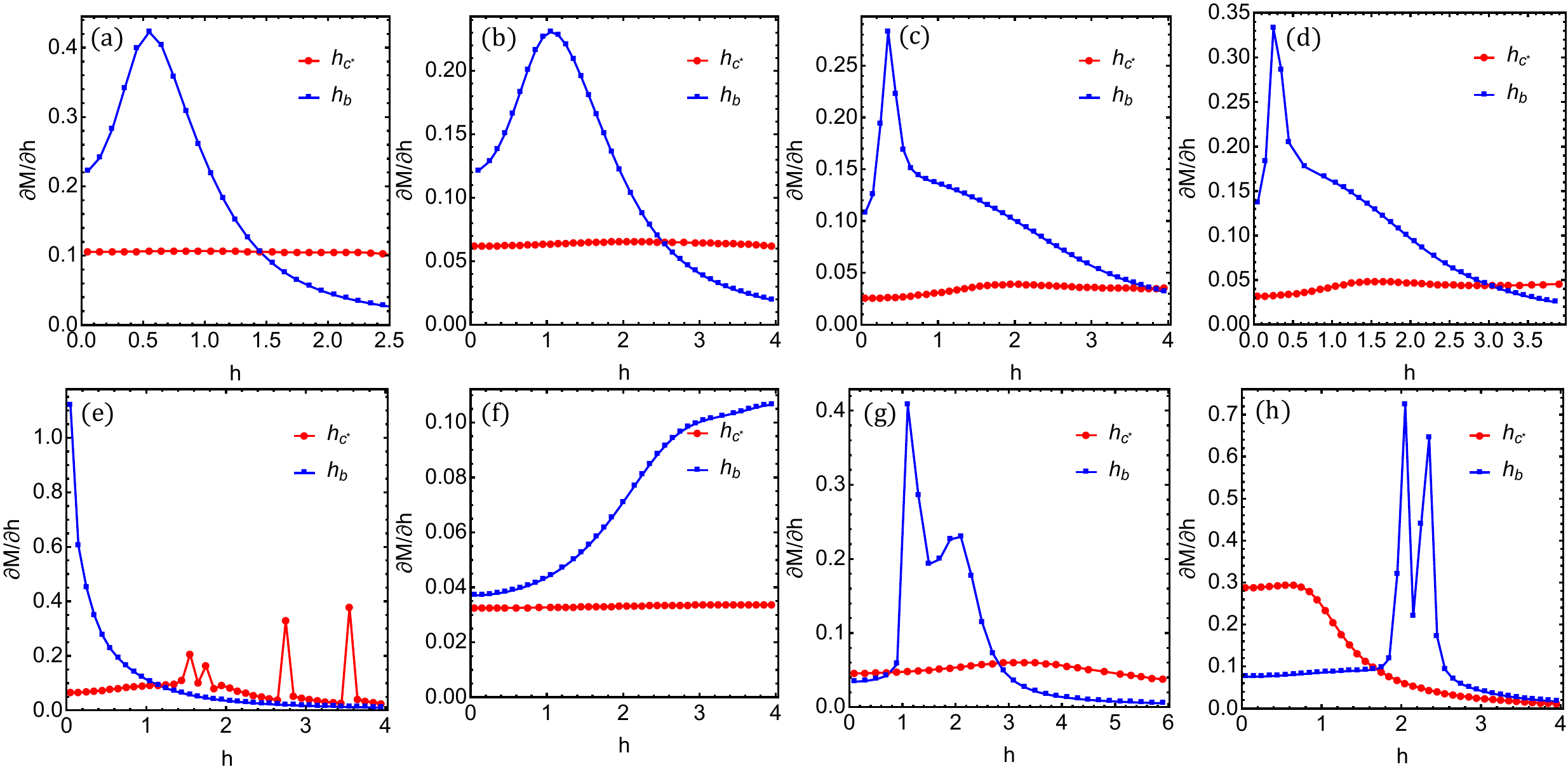}
\caption{The derivative of magnetization as a function of amplitude of field $h$. The results are obtained by ED calculation on 24-site PBC cluster.}
\label{Fig:dmag}
\end{figure}

\end{widetext}

\end{document}